# CEIMVEN: An Approach of Cutting Edge Implementation of Modified Versions of EfficientNet (V1-V2) Architecture for Breast Cancer Detection and Classification from Ultrasound Images

Sheekar Banerjee[1,2][0000-0003-1084-2351] and Md. Kamrul Hasan Monir[1]

[1] KaleidoSoft, Zagreb, Croatia
[2] IUBAT- International University of Business Agriculture and Technology, Dhaka 1230, Bangladesh
sheekar.banerjee@gmail.com

**Abstract.** Undoubtedly breast cancer identifies itself as one of the most widespread and terrifying cancers across the globe. Millions of women are getting affected each year from it. Breast cancer remains the major one for being the reason of largest number of demise of women. In the recent time of research, Medical Image Computing and Processing has been playing a significant role for detecting and classifying breast cancers from ultrasound images and mammograms, along with the celestial touch of deep neural networks. In this research, we focused mostly on our rigorous implementations and iterative result analysis of different cutting-edge modified versions of EfficientNet architectures namely EfficientNet-V1 (b0-b7) and EfficientNet-V2 (b0-b3) with ultrasound image, named as CEIMVEN. We utilized transfer learning approach here for using the pre-trained models of EfficientNet versions. We activated the hyper-parameter tuning procedures, added fully connected layers, discarded the unprecedented outliers and recorded the accuracy results from our custom modified EfficientNet architectures. Our deep learning model training approach was related to both identifying the cancer affected areas with region of interest (ROI) techniques and multiple classifications (benign, malignant and normal). The approximate testing accuracies we got from the modified versions of EfficientNet-V1 (b0- 99.15%, b1- 98.58%, b2- 98.43%, b3- 98.01%, b4- 98.86%, b5- 97.72%, b6- 97.72%, b7- 98.72%) and EfficientNet-V2 (b0- 99.29%, b1- 99.01%, b2- 98.72%, b3- 99.43%) are showing very bright future and strong potentials of deep learning approach for the successful detection and classification of breast cancers from the ultrasound images at a very early stage. The code for this research is available here: https://github.com/ac005sheekar/CEIMVEN- Breast.

## 1       Introduction

Deep learning techniques are being used in a wide range of industries and have demonstrated impressive computational performance in areas such as speech recognition, natural language processing, video analysis, image processing, computation, analysis and classification. Additionally, deep neural networks combine many layer types, including FC (fully connected) layers, convolutional neural networks (CNN), and recurrent neural networks (RNN) [1]. Multi-level deep neural networks (DNNs) are used to build diverse neural networks that can recognize and categorize traits from an input of huge unlabeled training data. CNNs appear to be the neural model that processes and analyzes images the best. We are getting closer at extracting features from photographs that encompass the many elements of the underlying problem as a result of CNN's performance. Since DCNN's convolution and subsampling layers may automatically extract visual features from a given patch, they do not require data-focused and appropriately decomposed cores. Big data processing demands a sophisticated infrastructure and combination of powerful devices. The processing of enormous amounts of data has been made easier by advancements in hardware processing units, which have also sped up research on DCNN. A graphics processing unit (GPU) with CUDA supportperforms applications 10 to 100 times faster than a central processor unit (CPU) with hundreds of compute cores. By using parallel computing, the typical classification difficulties for medical images can be effectively decreased. As a result, the machine learning approach needs simultaneous Deep Neural Network models using CPU and GPU.

Breast Cancer appears to be as most prevalent cancer in the world for women. The World Health Organization (WHO) reports that the leading causes of mortality worldwide in 2018 were severe diseases like heart disease and breast cancer. According to statistics provided by WHO, 17.7 million people worldwide lose their lives to cardiovascular illnesses (CVDs) each year, accounting for 31% of all fatalities worldwide. Heart attacks and strokes near to the end of their account for 80% of deaths [2].

It is a known fact that compared to other types of cancer, breast cancer has a disproportionately high mortality rate in both developing and wealthy nations. Although the causes of breast cancer remain mostly unknown, extensive research and investigation have shown that the disease occurs with some frequency. Hereditary family history, obesity or inadequate exercise, alcohol and tobacco addiction, exposure to ionizing radiation, having children too late or not at all, and not producing enough milk to feed the young are high-risk factors. The woman will experience symptoms of this disease as a result of these factors. For the treatment of breast cancer and increasing survival, early detection is crucial. Patients should receive an early diagnosis even when medical resources are limited in order to perform better. This can be done by being aware of early signs and symptoms and promptly referring them to therapy [3]. The risks of death as a result are going down because early-stage barriers are being effectively removed and access to medical services for breast cancer is being improved. Mammography, clinical biopsy, and routine manual self-check are just a few of the techniques that make up screening and have been studied as breast cancer assessment instruments. The most well-known of them is mammography, which records breast features using low-energy X-rays and generates the corresponding medical images for additional diagnosis. In comparison to other optional methods, mammography is thought to be one of the top contributions to lowering mortality and treating breast cancer because of its high efficacy. Cardio CTG using Doppler ultrasound monitoring, among other similar methods, is used for the early diagnosis of cardiac problems [4].

In our proposed research, we majorly concentrated on our exact implementations and iterative result analysis of several cutting-edge modified EfficientNet architectures, specifically EfficientNet-V1 (b0-b7) and EfficientNet-V2 (b0-b3) with ultrasound image. We named it as CEIMVEN (Cutting-Edge Implementation with Modified Efficient-Nets). Here, we used a transfer learning strategy to make use of the pre-trained EfficientNet versions models. In our specially modified EfficientNet architectures, we turned on the hyper-parameter tuning procedures, added completely linked layers, eliminated the rare outliers, and recorded the accuracy results. We used multiple classifications (benign, malignant and normal) and region of interest (ROI) algorithms to identify the cancer-affected areas in our deep learning model training strategy. The main goal of this research was to detect and classify breast cancer in real time from ultrasound images using more recent and cutting-edge deep neural network architectures. We sought to dispel the myth that improved accuracy may be obtained from using mammograms in medical image processing. We aimed to extract the best prediction accuracies from supposedly outdated ultrasound images by embedding cutting-edge neural networks into them. Unquestionably, we completed a really difficult and brave study project, and the outcomes were incredibly rich. The modification of EfficientNet-V1 (b0-b7) and EfficientNet-V2 (b0-b3) using a unique DL model and ultrasound images has never been attempted in prior studies.

Following the rest of this paper is: a literature review describing the previous research in the field of breast cancer detection and classification using other neural architectures, computational techniques, algorithms, datasets, optimizers; our proposed methodology of CEIMVEN implementation; experimental result analysis of our method. Finally, the conclusion and future research aspects originated from our research effort has been spotlighted.

## 2      Related Works

The medical imaging community has been conducting frequent observations in recent years thanks to the development of effective and potent CNN architectures. We reviewed the most related previous research works which we deeply admired but at the same time we implied to make improvements of too.

Experiments were conducted as a computer-generated diagnosis method for classification relying on both binary classification and multi-classification. They used a combined approach of neural networks (DNNs) like Res-Net-18, Shuffle-Net and Inception-Net-V3 by approaching transfer learning on the publicly available dataset of "BreakHis". For Res-Net, Inception-Net-V3, and Shuffle-Net, respectively, their method produced the best average accuracy for binary classification of benign or malignant cancer cases of 99.7%, 97.66%, and 96.94%. ResNet, Inception-Net-V3, and Shuffle-Net each had multi-class classification accuracy averages of 97.81%, 96.07%, and 95.79%, respectively [5]. A new multi-layered CNN architecture had been introduced for clinical mammograms with better cross-folding and accuracy more than 95% [6].

A synthesized image modelling using GAN-CNN-wavelet architecture with MIAS dataset has shown quite a good amount of accuracy (approximately 87%) previously [7]. In the study of "MultiNet", the authors showed an implementation of a concept related to transfer. The image analysis were conducted by using pre-trained models, including VGG16, DenseNet-201 and NasNet-Mobile, where their estimation of accuracy was 98% [8]. Four mammography imaging datasets—normal, benign, and malignant—were employed in a study, with the basic classifiers being a variety of deep CNN models, including Inception V4, ResNet-164, VGG-11, and DenseNet121. The Gompertz function was used to provide fuzzy rankings of the fundamental classification procedures as part of their ensemble approach. Particularly noteworthy was the 99.32% accuracy rate of their Inception V4 ensemble model using fuzzy rank-based Gompertz function [9]. In order to prepare computed tomography (CT), a study has proposed an automated segmentation model based on traditional CNN-based deep neural networks for the breast cancer CTV. Their model had three steps that functioned in a cascade, making it applicable to actual situations with roughly 80% accuracy [10]. By developing a multi-input classification model that took advantage of convolutional neural networks' strengths in image analysis, breast cancer detection using thermal images from various angles was made possible. The Database for Research with Infrared Image, the most extensively known public database of breast thermal pictures, was utilised in their application of our technique. Their top model had an accuracy of 97%, a ROC curve area of 0.99, and an 83% sensitivity [11]. Using an amassed dataset, a deep learning method was introduced for categorizing hematoxylin-eosin-stained breast cancer microscopy images into normal tissue, benign lesion, in situ carcinoma, and invasive carcinoma. This method utilized the Xception's six intermediate layers. This model had a kappa value of 0.965 and a precision of 97.79%. Additionally, it achieved a mean AUC-PR value of 0.991 and an average AUC-ROC score of 0.997 [12].

As a pre-processing step, the OMLTS-DLCN model used an adaptive fuzzy based median filtering (AFF) technique with the Shell Game Optimization (SGO) algorithm, which had a higher accuracy of 98.50% and 97.55% on the Mini-MIAS dataset and DDSM dataset, respectively [13]. A CNN model called CoroNet was suggested for use in the automatic detection of Breast Cancer using the CBIS-DDSM dataset. It used the Xception architecture, which had an overall accuracy of 94.92% after being pre-trained on the Image-Net [14]. Using a dataset provided by UCI and an F1 score of greater than 98, a study was conducted with the goal of developing a deep neural network that could predict the malignancy of breast cancer [15]. The categorization capacity of breast tumors applied to ultrasound pictures was seen in a study taking into account eight different fine-tuned pre-trained models. With regard to several performance measures, they developed a shallow bespoke convolutional neural network that performs better than the pre-trained models. In contrast to the best pre-trained model, which displays 92% accuracy and 0.972 AUC score, they estimated to attain 100% accuracy and achieve 1.0 AUC score [16].

A study using the segmentation model ResU-segNet and the hierarchical fuzzy classifier HFC, which combined the interval type-2 probabilistic fuzzy c-means (IT2PFCM) and fuzzy neural network (FNN), revealed a relatively low accuracy of 91% [17]. A development of two techniques for visualizing important nuclei in classification of histology images using graph convolutional networks had been shown with a range of accuracies between 92% and 94% [18]. An efficient classification model with a high performance up to 99.05% accuracy value had been researched using hyper-parameter optimization with parallel computing architecture and CUDA-enabled graphics processing unit [19]. An implementation of pooling structures had been applied on most CNN-based models, which might greatly improve the models' performance on mammographic image data with the same input, showed their overall accuracy close to 85% [20]. An implementation of deep learning algorithms using a conventional CNN and a recurrent CNN to differentiate three breast cancer molecular subtypes on MRI, had been shown where mean accuracy was close to 91% [21]. Again there were quite similar kind of implementations of CNNs, Res-Net50 and other neural networks related DL models by which they range of accuracies close to 85% to 90% [22-26].

Accordingly, we proposed a new convolutional neural network model as CEIMVEN with the implementation of modified Efficient-Net-V1 (b0-b7) and Efficient-Net-V2 (b0-b3). Our method came across with a far better overall accuracy than most other previous studies could possibly pull off. The descriptive representations about the accuracy and total result analysis can be found at the section of "Results and Accuracies".

## 3    Dataset and Proposed Method

For the extensive amount of data processing, We had to search through the database of live, open-source ultrasound images, collected from 600 female patients in 2018, aged between 25 and 75 years old [27]. Exactly 780 photos with an average size of 500 × 500 pixels made up the collection. Nevertheless, we executed a full-on data augmentation procedure for having almost 4500 more images from it for better training accuracies. All of the images were formatted into PNG whereas the ground truth images are presented with original images. Those images were categorized into 3 classes, which were identified as benign, malignant and normal; demonstrated in Fig.1.

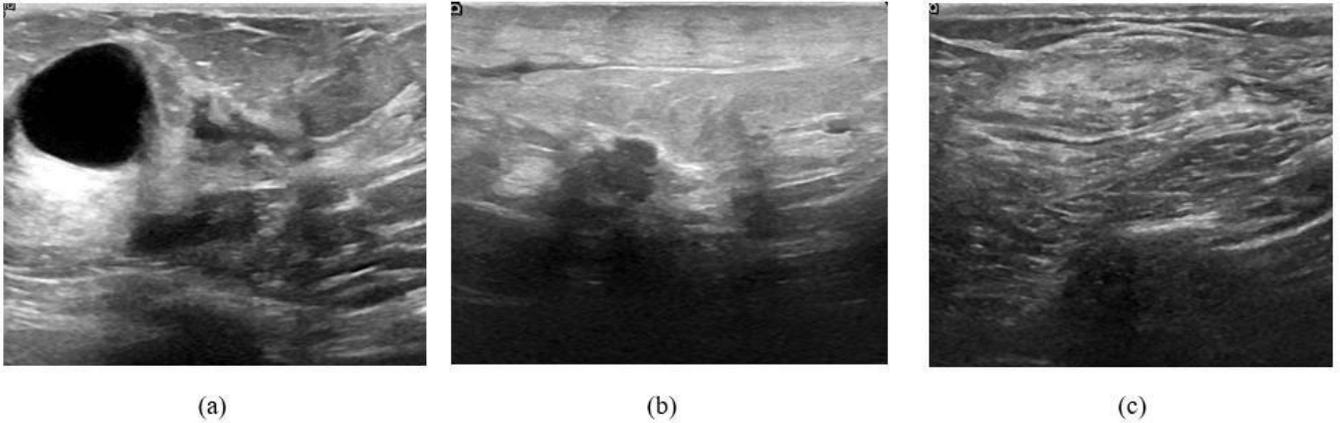

**Fig. 1.** Classified Ultrasound Image Dataset with 3 classes, identifying as stages of cancer; (a) benign (b) malignant and (c) normal.

Our proposed CEIMVEN method is detailed within two main components and they are: training the breast cancer detection and classification model with modified (a) EfficientNet-V1 with all of its sub-components (b0, b1, b2, b3, b4, b5, b6 and b7) and (b) the cutting-edge EfficientNet-V2 with all of its latest sub-components (b0, b1, b2 and b3).

At first, we had to fulfill and settle down with our complex requirement of computer vision related software and hardware machineries. For the deep learning and image processing environment we chose to utilize Pycharm IDE and train the deep learning (DL) model locally. To make our model compatible with using the same model on online platforms in the future, we saved our model as an H5 file and translated it to TensorFlow-JavaScript. We utilized CVAT for image data labeling with bounding boxes with the mask value we got from the open-source dataset for identifying the specific location of cancer cells from the ultrasound image. We were using the masking values as our reference point. Version 4.5.3.56 of OpenCV-Python and Python interpreter 3.9.1 with the majority of its global packages and variables were used. As a model backend, Keras in version 2.10.0 was employed. The main architectures of EfficientNet-V1 (b0-b7) and EfficientNet-V2 (b0-b3) were imported using Tensorflow version 2.6.1. We used an Intel Core i5- 10th Generation CPU with a clock speed of 2.11 GHz to be more precise with our gear. For the NVIDIA Geforce MX110 with 2GB Graphics Card, we installed CUDA version 10.2 and CUDA deep neural network libraries (CuDNN) for its largely compatible GPU support.

In Fig. 2, we manifested the implementation of our own neural network architecture with more improvisation and modification within conventional EfficientNet-V1 with 10 different layers. At first, we initiated an input layer of 3 × 3 filter size. Then we called the core pre-trained model of EfficientNet-V1. All of the sub-components such as: b0, b1, b2, b3, b4, b5, b6 and b7 were added and implemented gradually and iteratively with TensorFlow and Keras. We approached transfer learning methodology by using the pre-trained model as its core, fine-tuning it with more hyper-parameters and increasing its training and validation accuracies for a better research outcome.

After calling the core model, the input images moved from one convolution (Conv2D) layer with a 3 × 3 pixel entity filter size to another convolution (Conv2D) layer with a 3 × 3 pixel entity filter size. Conv2D's hidden layers started off with a filter size of 3 × 3 for the current preprocessing. Conv2D filter numbers were modified to 256 and 128 filters, respectively, in separate layers. In these parameters, the strides, padding, and activation functions were left unchanged. For effectively triggering multiple classification, we summoned the Rectified Linear Unit (ReLU) function. Then, using Conv2D, Dropout and Flatten layers as well as GlobalAveragePooling2D functions, the hyper-parameters were effectively optimized. For three distinct forms of stage detection (Benign, Malignant, and Normal), three (3) classes were established. We preprocessed and downsized the photos to a 224 x 224 pixel impression. Through the Flatten layer, the two dimensional tensor array values were transformed into one dimensional picture data. We added a Dropout layer as the 8$^{th}$ layer of the architecture. We had to initiate a dropout of 20% for keeping away the under-fitting scenarios of our DL model. We added another Dense layer having filter size of 3 × 3 for getting the two-dimensional array values into a single dimension.

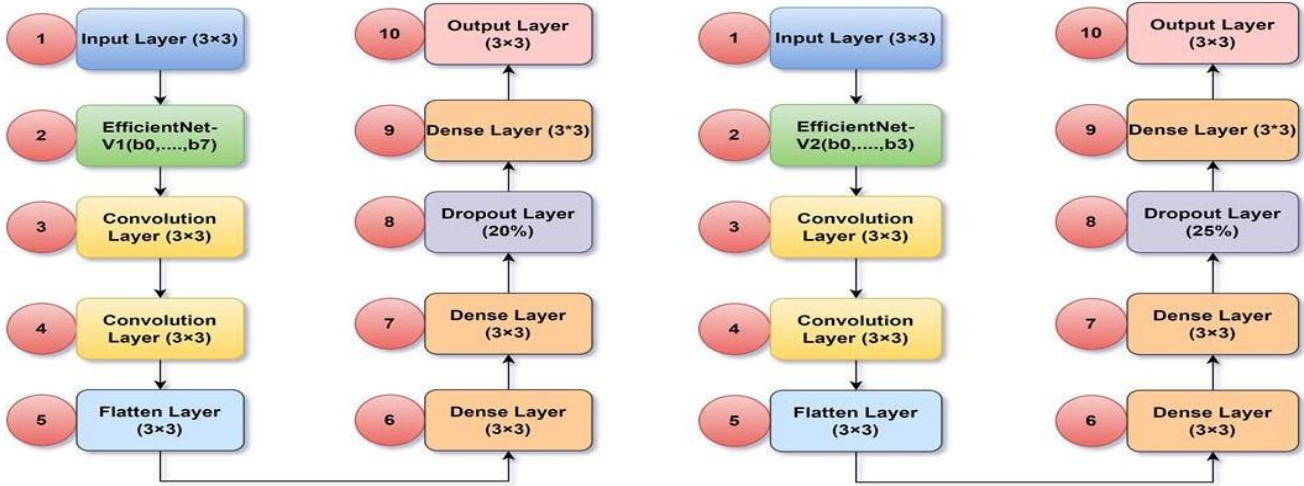

**Fig. 2.** The architecture of our proposed modified EfficientNet-V1 (b0-b7) and EfficientNet-V2 (b0-b3) model for cancer classification with 5% increase at V2 dropout (10 layers at each).

Lastly, we added an output layer with filter size of 3 × 3. 'Softmax' function was utilized for multi-class output. We could have used 'Sigmoid' function if we needed a binary classification only (Benign and Malignant). But as we were approaching for 3 classes, we used 'Softmax'. After we processed image data augmentation, we had to deal to a larger amount of image data. A significant number of epoch were also required for parameter adjustment, training, and testing on the large size of picture data. Device-installed GPU handled larger model training more quickly, although TensorFlow-GPU was also deployed as a backup. The batch size was 42, the epoch number was 500, and the initial learning rate was 0.01. To define vanishing gradient properties, stochastic gradient descent (SGD) was turned on. We used 4,100 picture samples for training, 200 for validation, and 200 for testing. We trained and saved the improvised EfficientNet-V1 (b0-b7) models as .h5 files and .pkl artifacts for further training and testing.

Quite similarly in the right part of Fig. 2, we demonstrated another implementation of our own neural network architecture with more improvisation and modification within conventional EfficientNet-V2 with another 10 different layers. At first, we initiated an input layer with filter size of 3 × 3. Then we called the core pre-trained model of EfficientNet-V2. All of the latest and cutting-edge sub-components such as: b0, b1, b2 and b3 were added and implemented gradually and iteratively with TensorFlow and Keras. We approached transfer learning methodology by using the pre-trained model as its core, fine-tuning it with more hyper-parameters and increasing its training and validation accuracies. After calling the core model, the input images moved from one convolution (Conv2D) layer with 3 × 3 pixel entity filter size to another convolution (Conv2D) layer with a 3 × 3 pixel entity filter size. Conv2D's hidden layers started off with filter size of 3 × 3 for the current preprocessing. Conv2D filter numbers were modified to 256 and 128 filters, respectively, in separate layers. In these parameters, the strides, padding, and activation functions were left unchanged. For effectively triggering multiple classification, the Rectified Linear Unit (ReLU) function was called. Then, using Conv2D, Dropout and Flatten layers as well as GlobalAveragePooling2D functions, the hyper-parameters were effectively optimized. For three distinct forms of stage detection (Benign, Malignant, and Normal), three (3) classes were established. We preprocessed and downsized the photos to a 224 x 224 pixel impression. Through the Flatten layer, the 2D tensor values were transformed into 1D picture data. We added a Dropout layer as the 8$^{th}$ layer of the architecture. We had to initiate a dropout of 25% for keeping away the under-fitting scenarios of our DL model which was significantly 5% higher than the previous EfficientNet-V1. We added another Dense layer with filter size of 3 × 3 for getting the two dimensional array values into a single dimension. Lastly, we added an output layer with filter size of 3 × 3. 'Softmax' function was utilized for multi-class output. After we processed image data augmentation, we had to deal to a larger amount of image data. A significant number of epoch were also required for parameter adjustment, training, and testing on the large size of picture data. Device-installed GPU handled larger model training more quickly, although TensorFlow-GPU was also deployed as a backup. The batch size was 42, the epoch number was 500, and the initial learning rate was 0.01. To define vanishing gradient properties, stochastic gradient descent (SGD) was turned on. We used 4,100 picture samples for training, 200 for validation, and 200 for testing. We trained and saved the improvised EfficientNet-V1 (b0-b7) models as .h5 files and .pkl artifacts for further training and testing.

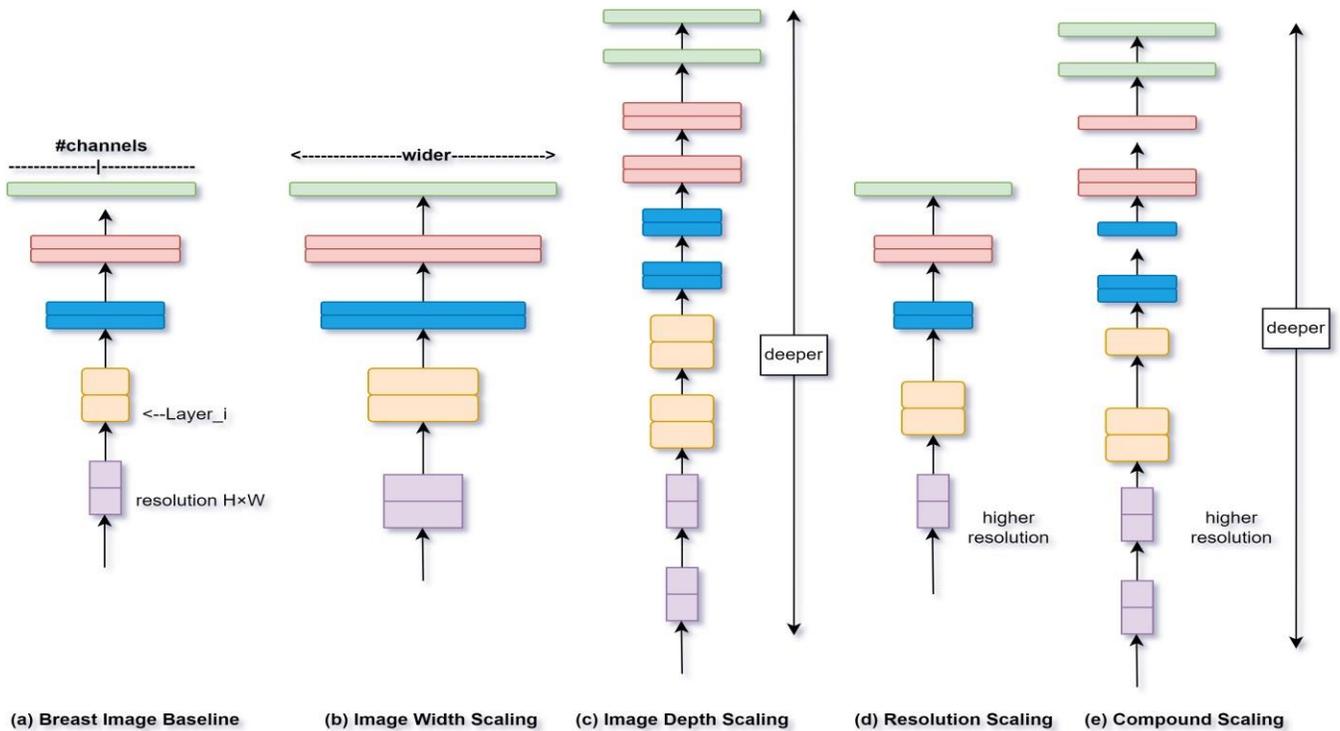

**Fig. 3.** EfficientNet Model scaling for our proposed method.

From Fig. 3 we can observe that (a) is a baseline network example for the breast ultrasound images. From (b) to (d) are conventional scaling that only increases one dimension of our neural network's width, depth and resolution. Therefore, (e) is our proposed compound neural scaling method which executes the uniform scaling of all three dimensions within a fixed aspect ratio. This is exactly what the EfficientNet-V1-V2 model stands for, showing it unique amalgamation of compound scaling in different scaling aspects and scenarios.

## 4     Results and Accuracies

As we entered into our CEIMVEN model testing phase, we divided our tasks into several steps such as: (a) recording the Area Under ROC Curve (AUC) of training and validation, (b) recording loss values of training and validation and (c) taking steps to improve the result values coming from the DL models.

**Table 1.** Modified EfficientNet-V1 models' average AUC (training, validation) for proposed breast cancer classification:

| Model | Training Accuracy (Approx.) | Validation Accuracy (Approx.) |
| --- | --- | --- |

| Model | Training Accuracy | Validation Accuracy |
|---|---|---|
| EfficientNet-V1-b0 | 99.45% | 82.47% |
| EfficientNet-V1-b1 | 99.21% | 82.73% |
| EfficientNet-V1-b2 | 99.29% | 83.89% |
| EfficientNet-V1-b3 | 99.67% | 82.93% |
| EfficientNet-V1-b4 | 99.85% | 84.31% |
| EfficientNet-V1-b5 | 99.79% | 84.86% |
| EfficientNet-V1-b6 | 99.83% | 85.27% |
| EfficientNet-V1-b7 | 99.89% | 85.68% |

In Table 1, we represented the average AUC in training and validation while we executed the DL model training with EfficientNet-V1 and its sub-components. Here, we can see the training accuracies of EfficientNet-V1's b0 to b7 fluctuates between different decimal values of 99%. Notably the training accuracy rose very negligibly higher from b0 to b7. While validation accuracies fluctuates between the values of 82.47% to 85.68%, signifies a good sign of accuracies than other deep neural nets like Inception-Net-V2 or ResNet-V1-V2.

**Table 2.** Modified EfficientNet-V1 models' average Loss (training, validation) for proposed breast cancer classification:

| Model | Training Loss (Approx.) | Validation Loss (Approx.) |
|---|---|---|
| EfficientNet-V1-b0 | 0.035 | 0.582 |
| EfficientNet-V1-b1 | 0.022 | 0.590 |
| EfficientNet-V1-b2 | 0.028 | 0.511 |
| EfficientNet-V1-b3 | 0.032 | 0.573 |
| EfficientNet-V1-b4 | 0.029 | 0.582 |
| EfficientNet-V1-b5 | 0.033 | 0.590 |
| EfficientNet-V1-b6 | 0.028 | 0.586 |
| EfficientNet-V1-b7 | 0.030 | 0.594 |

In Table 2, we represented the average loss in training and validation while we executed the DL model training with EfficientNet-V1 and its sub-components. Here, we can see the training losses of EfficientNet-V1's b0 to b7 fluctuates between the values of 0.022 to 0.035. Notably the training losses fluctuated quite inconsistently between b0 and b7. While validation losses fluctuated between the values of 0.511 to 0.594, signifies quite a good gap of losses. Simultaneously the losses shows significantly lower in numbers than other deep neural nets like Inception-Net-V2, VGG16-19 or ResNet-V1-V2.

**Table 3.** Modified EfficientNet-V2 models' average AUC (training, validation) for proposed breast cancer classification:

| Model | Training Accuracy (Approx.) | Validation Accuracy (Approx.) |
|---|---|---|
| EfficientNet-V2-b0 | 99.45% | 82.47% |
| EfficientNet-V2-b1 | 99.21% | 83.73% |
| EfficientNet-V2-b2 | 99.49% | 84.89% |
| EfficientNet-V2-b3 | 99.87% | 85.43% |

In Table 3, we represented the average AUC in training and validation while we executed the DL model training with EfficientNet-V2 and its sub-components. Here, we can see the training accuracies of EfficientNet-V2's b0 to b3 fluctuates between different decimal values of 99%, almost same as its predecessor V1. Notably the training accuracy rose very negligibly higher from b0 to b3. While validation accuracies fluctuates between the values of 82.47% to 85.43%, signifies a good sign of accuracies than other deep neural nets like VGG16-19, Inception-Net-V2 or ResNet-V1-V2.

**Table 4.** Modified EfficientNet-V2 models' average Loss (training, validation) for proposed breast cancer classification:

| Model | Training Loss (Approx.) | Validation Loss (Approx.) |
|---|---|---|
| EfficientNet-V2-b0 | 0.028 | 0.582 |
| EfficientNet-V2-b1 | 0.032 | 0.590 |
| EfficientNet-V2-b2 | 0.029 | 0.586 |

| EfficientNet-V2-b3 | 0.033 | 0.594 |

In Table 4, we represented the average loss in training and validation while we executed the DL model training with EfficientNet-V2 and its sub-components. Here, we can see the training losses of EfficientNet-V2's b0 to b3 fluctuates between the values of 0.028 to 0.033, almost same as its predecessor V1. Notably the training losses fluctuated quite less inconsistently between b0 and b3. While validation losses fluctuated between the values of 0.582 to 0.594, signifies quite a less gap of losses. Simultaneously the losses shows significantly lower in numbers than other deep neural nets like Mobile-Net, Inception-Net-V2 or ResNet-V1-V2.

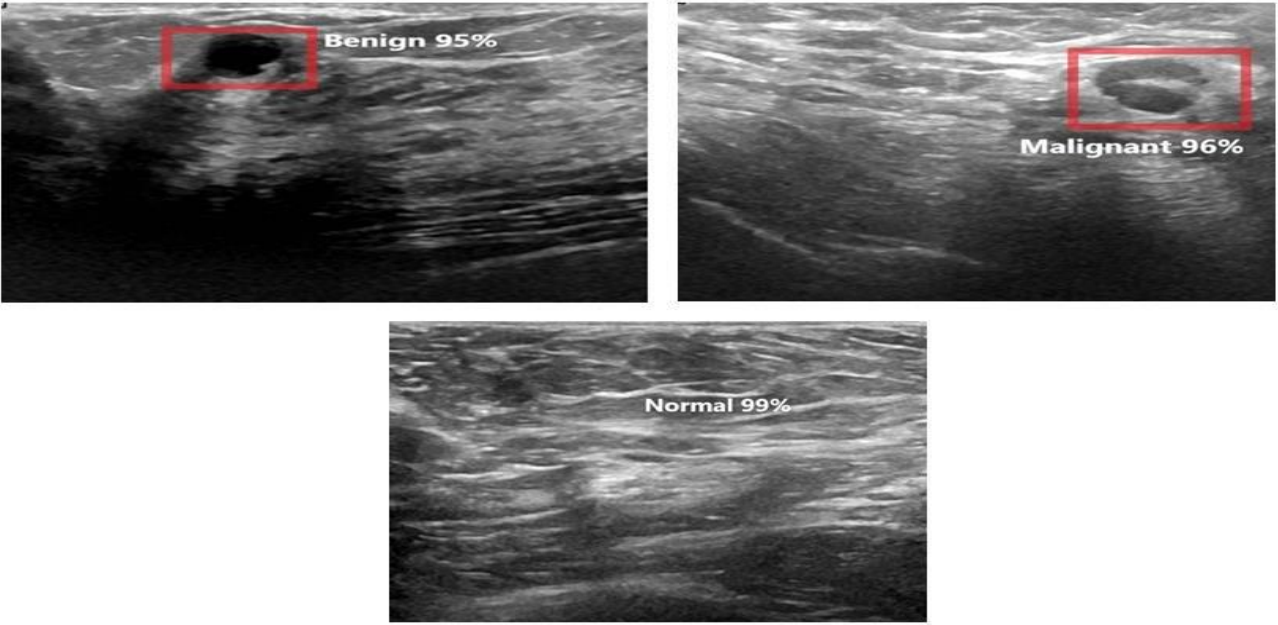

**Fig. 4.** Breast cancer cell detection from ultrasound image as "Benign" with 95% confidence, "Malignant" with 96% confidence and "Normal" with 99% confidence.

As we labelled the ultrasound breast images with CVAT upon its masked reference values, it helped us training another channel of the DL model as an object detector. Here, we used the cancer affected areas as the object which we wanted to detect and detect the class of the cancer simultaneously. We used the same image resolution for training and testing with a shape of $224 \times 224$ pixels. We tried to maintain the input layer of same $3 \times 3$ filter size and a single output layer with filter size of $3 \times 3$. We executed the input of the images one-by-one and got the prediction results with rectangular bounding boxes, identifying the cancer affected areas. The prediction results have been shown in Fig. 4.

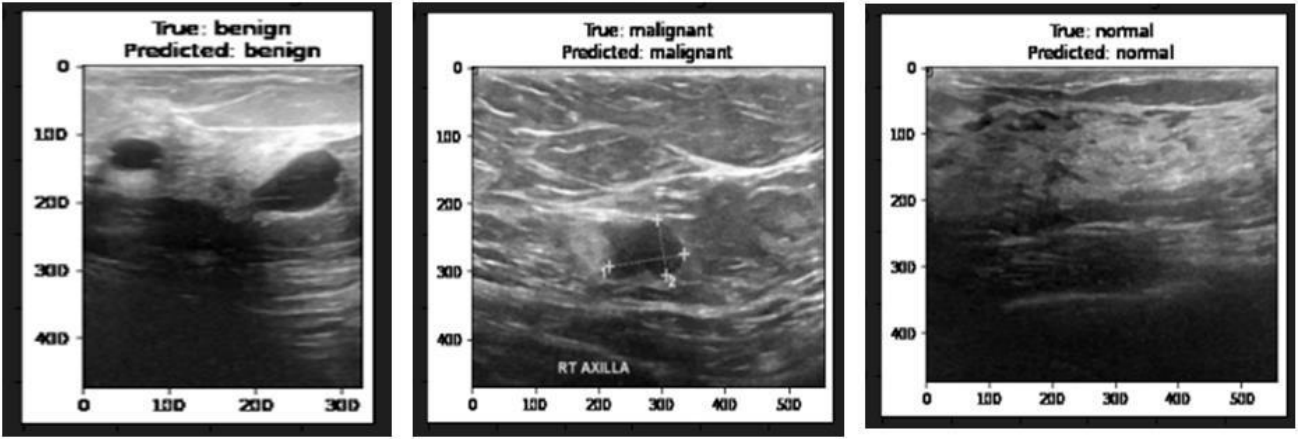

**Fig. 5.** Testing the prediction of the classes (benign, malignant and normal) from ultrasound images without identifying any cancer cell.

In Fig. 5, we demonstrated the classification and prediction capability of our implemented breast cancer DL model with the comparison from "true" to "predicted" landmark with a 2D graphical visualization. Here, we emphasized more on the cancer classification part rather than the detection of the cancer cell and affected breast areas. Combining all of the test cases the average prediction accuracy was approximately 97.79%. We dealt with a generative confusion matrix where concerned about the four important factors like: True Positive, True Negative, False Positive and False Negative. The average precision rate was approximately 1.36% while the recall rate was approximately 0.84%. All of these detection and prediction result factors for each classes are demonstrated in Table 5.

**Table 5.** Accuracies of Prediction and Detection, Precision and Recall rate for each classes:

| Model | Prediction Accuracy (Approx.) | Precision Rate (Approx.) | Recall Rate (Approx.) |
| --- | --- | --- | --- |
| Benign | 97.86% | 1.89% | 0.25% |
| Malignant | 96.27% | 1.95% | 1.78% |
| Normal | 99.26% | 0.25% | 0.49% |

At first, the average training accuracy for affected area detection was over 97%, while the average testing accuracy was over 91.47%. With such parameters as "re-scale" = 1./255, "horizontal-vertical flip", "rotation range" = 360, "width-height" shift, and "zoom-brightness" range with its important aperture, we attempted to recreate a number of picture data augmentation functions. We managed to renew the image data and materially improve the dataset. We used the same iteration approach and epoch numbers to retrain the model. The retesting procedure was started by us. The testing accuracy thereafter had a huge increase, ending at 96.49% on an average landmark. We noticed much greater accuracy when we increased the batch size to 54 and the epoch number to 600. Overall, the CNN-algorithmic improvised EfficientNet-V1-V2 based ultrasound image data model training and prediction testing performed with an average accuracy rate of 97.79% allowed us to carry out the integrated process of detection and prediction.

## 5 Conclusion and Future Endeavors

Our main focus in this study was the real-time identification and classification of breast cancer from ultrasound pictures using more recent and state-of-the-art deep neural network designs. We wanted to break the stereotype of using mammograms from the medical image analysis and expecting better accuracies from that. We wanted to embed state of the art neural models into so called obsolete ultrasound images and drag the best prediction accuracies out of it. Undoubtedly we pulled off a very bold and challenging research work and significantly rich results out of it. No previous research work has ever tried the modification of EfficientNet-V1 (b0-b7) and EfficientNet-V2 (b0-b3) with a custom DL model along ultrasound images. That's why we approached this method for the creation of a unique research austerity and achievement of a significantly higher overall accuracy (97.79%). With overall data augmentation we used a total number of 5280 images while we partitioned the entire dataset into (train, validation, test) portions as (80%, 10%, 10%) channels. 10% of the testing data were precisely used one-by-one as input for getting the prediction results.

In the future, we will strive to collaborate with more robust datasets from various reputed cancer institutes in the world.
We are confident that adding more patient image datasets would improve the capabilities and prediction accuracy of our DL artifacts. In order to improve the diagnosis of breast cancer with higher accuracies while taking into account cutting-edge techniques like Inception-ResNet-V2, Fast-AI, Dense-Net (121, 169, 201) and so on, we have significant plans to expand our vigorous research work.